\documentclass[prl,twocolumn,showpacs]{revtex4}
\usepackage[french]{babel}
\usepackage{amsmath}
\usepackage{amssymb}
\usepackage[T1]{fontenc}
\usepackage{graphicx}

\begin{document}

\title{Statistical Aging and Non Ergodicity\\
in the Fluorescence of single Nanocrystals}
\author{X. Brokmann, J.-P. Hermier*, G. Messin, P. Desbiolles, J.-P. Bouchaud$^{\dagger}$ and M. Dahan}

\affiliation{Laboratoire Kastler Brossel, École Normale Supérieure et Université
Pierre et Marie Curie, 24, rue Lhomond, 75231 Paris Cedex 05, France. \\
*Pôle Matériaux et Phénomènes Quantiques, Université Denis Diderot, 2, place Jussieu, 75251 Paris Cedex 05, France.\\
$^{\dagger}$Service de Physique de l'Etat Condensé, Commissariat à l'Énergie Atomique, Orme des Merisiers, 91191
Gif-sur-Yvette Cedex, France. }

\begin{abstract}
The relation between single particle and ensemble measurements is adressed for semiconductor CdSe nanocrystals. We record their fluorescence at the single-molecule level and analyse their emission intermittency, which is governed by unusual random processes known as Lévy statistics. We report the observation of statistical aging and ergodicity breaking, both related to the occurrence of Lévy statistics. Our results show that the behaviour of ensemble quantities, such as the total fluorescence of an ensemble of nanocrystals, can differ from the time averaged individual quantities, and must be interpreted with care.
\end{abstract}

\pacs{05.40.Fb, 78.67.Bf}

\maketitle  The relation between single particle and ensemble measurements is at the core of statistical physics, and becomes
crucial given that experiments are now able to resolve individual nanometer-sized objects. This question is addressed
here for semiconductor CdSe nanocrystals. The fluorescence properties of these colloidal quantum dots (QDs) have raised great attention due to their size-induced spectral tunability, high quantum yield and remarkable photostability at room temperature
\cite{Alavisatos96}, all of which make QDs a promising system for biological labelling \cite{Alavisatos98},
single-photon sources \cite{Michler00} and nanolasers \cite{Klimov00}. 

When studied at the single molecule level, CdSe QDs share with a large variety of other fluorescent nanometer-sized systems \cite{Moerner97,Dickson97,Barbara97} the
property of exhibiting fluorescence intermittency \cite{Nirmal96}. This means that the fluorescence intensity randomly
switches from bright ('On') states to dark ('Off') states under continuous excitation. Although the very origin of the
intermittency for CdSe QDs remains a matter of investigation, its statistical properties have been studied. For a given
QD, the durations $\tau_{\mathrm{on}}$ and $\tau_{\mathrm{off}}$ of the On and Off periods follow slowly decaying
power-law distributions $P_{\mathrm{on}}(\tau_{\mathrm{on}}>\tau)=\left(\tau_{0}/\tau\right)^{\mu_{\mathrm{on}}}$,
$P_{\mathrm{off}}(\tau_{\mathrm{off}}>\tau)=\left(\tau_{0}/\tau\right)^{\mu_{\mathrm{off}}}$, where $\mu_{\mathrm{on}}$
and $\mu_{\mathrm{off}}$ are close to 0.5 \cite{Kuno00,Kuno01,Shimizu01}. This behaviour extends over several orders of magnitude, from the detection integration time $\tau_{0}$ up to hundreds of seconds, with very small dependence on temperature or excitation intensity.

The crucial point for our analysis is that both $\mu_{\mathrm{on}}$ and $\mu_{\mathrm{off}}$ are smaller than 1.
In this case, the decay is so slow that the mean value of $P_{\mathrm{on}}$ and $P_{\mathrm{off}}$ is formally
infinite, and very long events tend to dominate the fluorescence signal, producing strong intermittency. The duration
of the On and Off periods are thus governed by "Lévy statistics", which have been encountered in various fields
\cite{Shlesinger95,Bardou01,Bouchaud90,Ott90,Klafter96,Zaslavsky99,Bouchaud95,Mandelbrot97,DaCosta00} such as laser
cooling of atoms \cite{Bardou01}, dynamics of disordered \cite{Bouchaud90} and chaotic \cite{Klafter96} systems, glassy
dynamics \cite{Bouchaud95} or economics and finance \cite{Mandelbrot97}.

\begin{figure}[h!]
\includegraphics*[width=4cm]{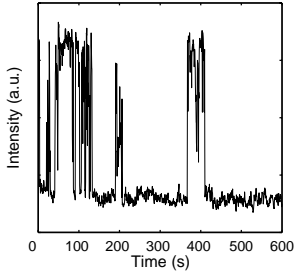}
\caption{Fluorescence intermittency of a single CdSe nanocrystal measured over 10 minutes with a 100 ms time bins. Due to the broad
distribution of the On and Off states, the signal is dominated by a few long events.} \label{fig1}
\end{figure}

In this Letter, we show that single QD measurements can be used to explicitly compare ensemble- and time-averaged
properties and explore some of the unusual phenomena induced by Lévy statistics, such as statistical aging and
ergodicity breaking. Using an epi-fluorescence microscopy set-up and a low-noise CCD camera, we simultaneously recorded
at room temperature the fluorescence intensity of 215 individual QDs for duration of 10 minutes with a time resolution
of 100 ms \cite{Preparation}. The blinking of the fluorescence intensity was observed for each QD detected in the field
of the camera (Fig.\ref{fig1}). Due to the binary behaviour of the blinking process, each intensity time trace was
simply considered as a sequence of $n$ On and Off times
$\{\tau_{\mathrm{on}}^{(1)},\tau_{\mathrm{off}}^{(1)},\tau_{\mathrm{on}}^{(2)},\tau_{\mathrm{off}}^{(2)},...,
\tau_{\mathrm{on}}^{(n)},\tau_{\mathrm{off}}^{(n)}\}$  from which the distributions $P_{\mathrm{on}}$ and $P_{\mathrm{off}}$ were derived. In our measurements, the On and Off periods both followed power-law distributions \cite{Trunc}. After adjustment of the cumulative distributions of the On and Off periods for each of the 215 QDs, the exponents $\mu_{\mathrm{on}}$ and $\mu_{\mathrm{off}}$  were estimated to be respectively 0.58 (0.17) and 0.48 (0.15), consistent with previous experiments \cite{Kuno01,Shimizu01}. For all pair of QDs, we also computed the Kolmogorov-Smirnov (KS) likelihood estimator \cite{Kolmogorov} to compare the On (resp. Off) distributions between each pair of QDs. For our set of data, the KS tests yield the same average value of 0.4 (0.3) for both On and Off distributions, well above the value 0.05 usually considered as an inferior limit to assume that two datasets have identical distributions. In the following, the 215 QDs are therefore considered as statistically identical, with $\mu_{\mathrm{on}}$=0.58 and $\mu_{\mathrm{off}}$=0.48 \cite{Kolmogorov2}.

\begin{figure}
\includegraphics[width=8cm]{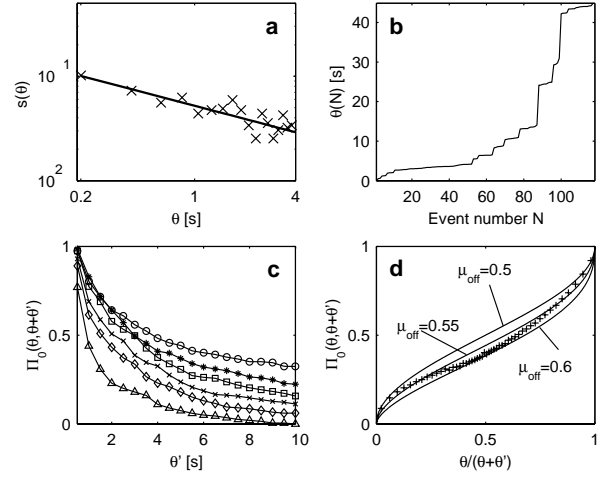}
\caption{Statistical aging measured from a sample of 215 QDs. (a) Logarithmic plot of the probability density
$s(\theta)$ for a QD to jump in the On state after having spent a total time $\theta$ in the Off state. The solid line
is a power-law adjustment $\theta^{-\alpha}$ with $\alpha = 0.5$. (b) Evolution of the total time spent in the Off
state $\theta(N)$ with the number $N$ of Off periods for a given QD ; the sum $\theta(N)$ is dominated by a few events
of the order of $\theta(N)$ itself. (c) Persistence probability $\Pi_{0}(\theta,\theta+\theta')$ measured from the set
of 215 QDs for $\theta = 0.1$ s ($\triangle$), $\theta= 0.5$ s ($\diamond$), $\theta = 1$ s ($\times$), $\theta = 2$ s
($\square$), $\theta = 4$ s ($\ast$), $\theta = 8$ s ($\circ$). $\Pi_{0}(\theta,\theta+\theta')$ depends on $\theta$,
indicating that the process is aging. (d) Persistence probability $\Pi_{0}(\theta,\theta+\theta')$ measured for
$\theta$ and $\theta'$ ranging between 0 and 10 s with 0.1 s time bins and expressed as a function of
$\theta/(\theta+\theta')$ (+). Each point corresponds to the average over 200 adjacent data points. The solid lines are
the theoretical predictions  for exponents $\mu_{\mathrm{off}} = 0.5$, $\mu_{\mathrm{off}}=0.55$ and
$\mu_{\mathrm{off}}= 0.6$.} \label{fig2}
\end{figure}

The first observation is that, for purely statistical reasons, the fluorescence of QDs is non-stationary, i.e. time
translation invariance is broken in the intermittency process. This is best evidenced by studying the rate at which the
QDs jump back from the Off to the On state (a "switch on" event). For this purpose, we computed the ensemble average of
the probability density $s(\theta)$ to observe a QD switching on between $\theta$ and $\theta + d\theta$ after a time
$\theta$ spent in the Off state. For Off periods following a "narrow" distribution (with a finite mean value
$\langle\tau_{\mathrm{off}}\rangle$), $s(\theta)$ - also called the renewal density - would be independent of $\theta$
and equal to $1/\langle\tau_{\mathrm{off}}\rangle$. The situation is drastically changed for CdSe QDs: due to the fact
that $\mu_{\mathrm{off}} <1$, $\langle\tau_{\mathrm{off}}\rangle$ no longer exists and $s(\theta)$ is then expected to
decay as $\theta^{-(1-\mu_{\mathrm{off}})}$ \cite{Bardou01}. This means that as time grows, the switch on events occur
less and less frequently. Fig.\ref{fig2}a shows that our data match these theoretical predictions: the measured value
of $s(\theta)$ decreases like $\theta^{-\alpha}$, with $\alpha = 0.5$ in agreement with the value $(1-\mu_{\mathrm{off}})$
expected from our measurement of $\mu_{\mathrm{off}}$.

This non-stationary behaviour can be understood by considering, for each QD, the quantity :$$\theta(N)=\sum_{i=1}^{N}
\tau_{\mathrm{off}}^{(i)},$$i.e. the total time spent in the Off state during the $N$ first Off periods
(Fig.\ref{fig2}b). Assuming that the $\tau_{\mathrm{off}}^{(i)}$ are independent and $\tau_{\mathrm{off}}$ having no
mean value, the sum of $N$ such independent random variables must be evaluated by means of the Generalized Central
Limit Theorem (see e.g. \cite{Bardou01}). This theorem states that $\theta(N)$, instead of scaling as $N$, grows more
rapidly, as $N^{1/\mu_{\mathrm{off}}}$. As shown in Fig.\ref{fig2}b, the sum $\theta(N)$ is dominated by few events.
This central property, distinctive of Lévy statistics, means that, as time grows, one observes long events that are of
the order of $\theta(N)$ itself \cite{NarrowMax}. Hence, the probability for a QD to switch on decreases with time: the
system ages \cite{Bouchaud95} and the signal is non-stationary.

To test the assumption that the Off events are independent and to gain further insight into this aging effect, we
computed the persistence probability $\Pi_{0}(\theta,\theta+\theta')$, defined as the probability that no switch on
event occurs between $\theta$ and $\theta+\theta'$. In the case of independent Off periods with an exponential
distribution (with mean value $\langle\tau_{\mathrm{off}}\rangle$), $\Pi_{0}(\theta,\theta+\theta')$ is independent of
$\theta$, and given by $e^{-\theta'/\langle\tau_{\mathrm{off}}\rangle}$, illustrating that the switching process is
stationary. The computation of $\Pi_{0}$ from our dataset reveals a completely different pattern: the probability that
no switch on event occurs within a given duration $\theta'$ decreases with $\theta$ (Fig.\ref{fig2}c), consistent with
the behaviour of $s(\theta)$. Furthermore, $\Pi_{0}(\theta,\theta+\theta')$ is found to depend only on the reduced
variable $\theta/(\theta+\theta')$, and to vanish for $\theta/(\theta+\theta')$ close to 0 (Fig.\ref{fig2}d). This
result proves that one has to wait a time $\theta'$ of the order of $\theta$ to have a chance to observe a switch on
event, in qualitative agreement with the fact that the largest term of the sum $\theta(N)$ is of the order of
$\theta(N)$ itself. Quantitatively, for independent Off events distributed according to a Lévy distribution
$P_{\mathrm{off}}$ with exponent $\mu_{\mathrm{off}}$ , the persistence probability is expected to
read:$$\Pi_{0}(\theta,\theta+\theta')=\int_{0}^{\theta/(\theta+\theta')}\beta_{\mu_{\mathrm{off}},1-\mu_{\mathrm{off}}}(u)du$$
where $\beta$ is the beta distribution on [0,1] \cite{Bouchaud95,Godreche01}. Our data follow this prediction with
$\mu_{\mathrm{off}}$ = 0.55, in agreement both with our previous estimations of $\mu_{\mathrm{off}}$ (Fig.\ref{fig2}d)
and with the assumption that the Off events are independent. These results show that the aging effect has a pure
statistical origin and is not related to an irreversible process (such as photo-destruction). Due to the statistical
properties of Lévy distributions, non-stationarity emerges despite the time-independence of the laws governing the
microscopic fluorescence process.

\begin{figure}
\includegraphics[width=8cm]{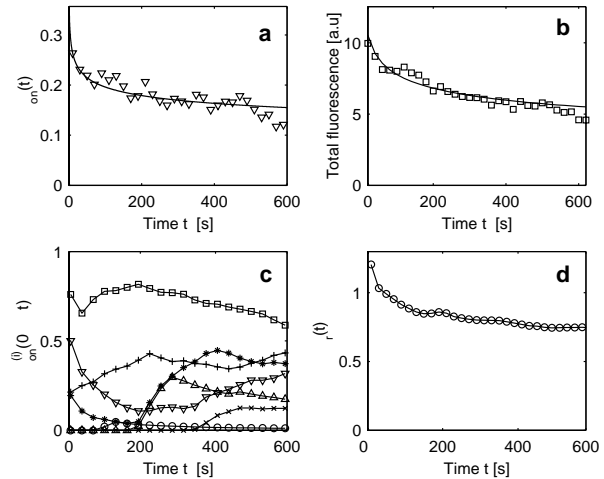}
\caption{Non-stationarity and non-ergodicity in a sample of QDs. (a) Time evolution of the fraction
$\Phi_{\mathrm{on}}(t)$ of QDs in the On state at time $t$ ($\nabla$). $\Phi_{\mathrm{on}}(t)$ decays as
$t^{-0.13}=t^{\mu_{\mathrm{off}}-\mu_{\mathrm{on}}}$ (solid line). (b) Time evolution of the total fluorescence signal
emitted by the sample ($\square$): the darkening effect follows a $t^{-0.18}$ power-law decay (solid line). (c) Typical
time evolution of  $\Phi_{\mathrm{on}}^{(i)}(0 \mapsto t)$ - the fraction of time spent in the On state between 0 and
$t$ - for 7 QDs. The time averages are widely fluctuating, even in the long integration time limit. (d) Evolution of
the relative dispersion $\sigma_{r}(t)$ of $\Phi_{\mathrm{on}}^{(i)}(0 \mapsto t)$ at time $t$ over the ensemble of QDs
($\circ$). As time grows, $\sigma_{r}(t)$ tends to a constant value, illustrating that the time averages trajectories do not
converge to any asymptotic value.} \label{fig3}
\end{figure}

From a more general standpoint, this non-stationary behaviour has also profound consequences on basic data
interpretation, such as the ensemble-averaged total fluorescence emitted by a population of QDs. We illustrated this by
studying $\Phi_{\mathrm{on}}(t)$, the fraction of QDs in the On state at a given time $t$ (Fig.\ref{fig3}a). In the
context of Lévy statistics, the time evolution of $\Phi_{\mathrm{on}}(t)$ is intimately linked to the relative amount
of time spent in the On and Off states for each QD. Qualitatively, the Off events tend to be dominant whenever
$\mu_{\mathrm{off}}<\mu_{\mathrm{off}}$ since $\theta(N)=\sum_{i=1}^{N}\tau_{\mathrm{off}}^{(i)}$  grows faster than
its counterpart $\widehat{\theta}(N)=\sum_{i=1}^{N}\tau_{\mathrm{on}}^{(i)}$ . When analysed in a more quantitative
way, the fraction $\Phi_{\mathrm{on}}(t)$ can be shown to decrease asymptotically as
$t^{\mu_{\mathrm{off}}-\mu_{\mathrm{on}}}$ \cite{Bardou01}. Experimental results confirm this analysis:
$\Phi_{\mathrm{on}}(t)$ decays as $t^{-\beta}$, with an exponent $\beta = 0.13$ indeed consistent with the previous
determination of $\mu_{\mathrm{on}}$ and $\mu_{\mathrm{off}}$ (Fig.\ref{fig3}a). We also observed that the average
signal over the whole CCD detector - i.e. the sum of the fluorescence of all the QDs - decays as $t^{-0.18}$, in
agreement (within experimental uncertainty) with the fact that time increasing, less and less QDs are in the On state,
causing the total fluorescence to decrease like $\Phi_{\mathrm{on}}(t)$ (Fig.\ref{fig3}b). Importantly, we also
observed that this fluorescence decay is laser-induced and reversible: after a continuous laser illumination of 10
minutes, leaving the sample in the dark for about 10-15 minutes systematically lead to a complete recovery of its initial fluorescence.
This confirms that this decay is again purely statistical, and not related to an irreversible bleaching of the QDs.

Our final observation focuses on non-ergodic aspects of random processes driven by Lévy statistics. Single particle
measurements allow one to compare directly $\Phi_{\mathrm{on}}(t)$ and the fraction of time $\Phi_{\mathrm{on}}^{(i)}(0
\mapsto t)$ spent in the On state between 0 and $t$ for the $i$th QD. This provides a direct test of the ergodicity of
the QD fluorescence. While the ensemble average $\Phi_{\mathrm{on}}(t)$ decays deterministically as $t^{-0.13}$
(Fig.\ref{fig3}a), each time average widely fluctuates over time and for a given $t$, the values of
$\Phi_{\mathrm{on}}^{(i)}$ are broadly distributed between 0 and 1, even after a long time of integration
(Fig.\ref{fig3}c). To study the behaviour of time averages, we calculated the relative dispersion $\sigma_{r}(t)$ of
the time averages at time $t$, where $\sigma_{r}(t)$ corresponds to the standard deviation of the distribution of
$\Phi_{\mathrm{on}}^{(i)}(0\mapsto t)$ over the set of QDs, divided by its mean value. Fig.\ref{fig3}d shows that
$\sigma_{r}(t)$ does not decay to zero, and is still of order 1 on the experimental time scale. Therefore, even for
long acquisition times, the fluctuations of the time averages from QD to QD remain of the order of the time averages
themselves and do not vanish as expected for ergodic systems. These data indicate ergodicity breaking: due to rare
events with a duration comparable to the total acquisition time, there is no characteristic time-scale over which
physical observables can be time-averaged. Even for long acquisition time, $\Phi_{\mathrm{on}}^{(i)}(0 \mapsto t)$ does
not converge and no information on the ensemble value $\Phi_{\mathrm{on}}$ can be obtained by time averaging an
individual trajectory.

While we found that accurate estimates of $\mu_{\mathrm{on}}$ and $\mu_{\mathrm{off}}$ are essential to analyze and
predict the statistical properties of the fluorescence, the microscopic origin of these broad distributions is not yet
established. Distributions of Off times are sometimes attributed to distributions of static traps from which the charge
of an ionised QD escapes by tunnelling effect \cite{Kuno01,Orrit02}. In these models, the value of $\mu_{\mathrm{off}}$
strongly depends on microscopic characteristics of the QDs, and it is not clear how this is compatible with the
statistical homogeneity of the different QDs suggested by the KS test. The dynamic changes of the particle environment
are also often invoked to account for the fluctuating emission of the QD \cite{Shimizu01,Silbey02}. Some authors have
thus suggested models in which the trap for the charge of the ionised QD follows a random walk in a 1D parameter space,
yielding a universal value 1/2 for $\mu_{\mathrm{off}}$ \cite{Shimizu01}. However both of these models (static and
dynamic) have yet to be more thoroughly tested. Since intermittency is an ubiquitous process at the nanometer scale,
some of the arguments discussed here for QDs might also apply to other systems. In particular, our analysis shows that
non-stationary behaviour of the fluorescence - sometimes attributed to photochemical processes - can also have purely
statistical origins (such as statistical aging). Recent evidence have shown that this may be the case in a system as
microscopically different from QDs as green fluorescent proteins\cite{Amblard01}. In this respect, aging and
non-ergodicity might be an important pattern when studying single nanometer-sized objects in complex environments.

In conclusion, our experimental results show that ensemble averaged fluorescence properties of individual CdSe QDs are deeply affected by the non-standard statistical properties of the Lévy statistics governing the blinking process. We found that a population of QDs exhibit statistical aging. Hence, despite the blinking statistics are time-independent, the fluorescence emitted by an ensemble of QDs under continuous laser excitation is non-stationary. Our data also evidence that due to the scaling properties of Lévy statistics, CdSe QDs are non-ergodic systems : time- and ensemble-averaged properties do not coincide anymore, in full contrast with usual assumptions when studying nanoscale emitters.

We acknowledge fruitful discussion with F. Amblard, C. Cohen-Tannoudji., C. Godrèche and J. M. Luck. We thank E.
Giacobino for valuable comments on the manuscript. This work was supported by CNRS. Correspondence and requests for materials should be addressed to M.D. (maxime.dahan@lkb.ens.fr).


\begin{thebibliography}{99}
\bibitem{Preparation}{We spin coated a nanomolar solution of CdSe/ZnS QDs (1.8 nm core radius, 570 nm peak emission)
on a glass coverslip, covered by a thin film of PMMA. The nanoparticules were excited in epifluorescence using the 488
nm line of an Ar$^{+}$ laser at an intensity of 0.1 kW/cm$^{2}$. Their fluorescence was recorded using a CCD camera
(Cool-Snap, Roper Scientific). Using antibunching measurements \cite{Michler00,Messin01} we had previously checked
that, for these conditions of sample preparation, the surface concentration of the nanoparticules was low enough so
that they could be detected individually.}
\bibitem{Trunc}{In agreement with previous measurements \cite{Shimizu01}, we also observed that
the On periods display a truncation time $\tau_{\mathrm{trunc}}$ after which $P_{\mathrm{on}}$ would fall more rapidly.
However, for our experimental conditions, $\tau_{\mathrm{trunc}}$ is greater than tens of seconds, fluctuating from QD to QD. On the
time scale of our analysis, the occurrence of this truncation does not affect our results.}
\bibitem{Kolmogorov2}{All the results shown in this Letter are found unchanged when working with a subset of QDs
for which the average values of the KS test are as high as 0.62 (0.3) for $P_{\mathrm{on}}$ and 0.63 (0.30) for
$P_{\mathrm{off}}$, confirming the statistical homogeneity of the sample.}
\bibitem{NarrowMax}{This result is in striking contrast with the case of narrow distributions. For an exponential distribution $ae^{-at}$, the largest
event is only of the order of $\log(N)/a$).}
\end{thebibliography}
\end{document}